\documentclass{article}

\usepackage{graphicx}
\usepackage{psfig}
\usepackage{epsfig}
\usepackage[round]{natbib}

\title{
Probabilistic forecasting of temperature: comments on the
Bayesian Model Averaging approach
}

\author{Stephen Jewson\footnote{\emph{Correspondence address}: RMS, 10 Eastcheap,
London, EC3M 1AJ, UK. Email: \texttt{x@stephenjewson.com}}}
\begin{document}

\maketitle

\begin{abstract}
A specific implementation of Bayesian model averaging has recently
been suggested as a method for the calibration of ensemble
temperature forecasts. We point out the similarities between this
new approach and an earlier method known as kernel regression. We
also argue that the Bayesian model averaging method (as applied)
has a number of flaws that would result in forecasts with
suboptimally calibrated mean and uncertainty.
\end{abstract}

\section{Introduction}

There is significant demand within industry for adequate
probabilistic forecasts of temperature. However, this demand has
not been met by the meteorological community and such forecasts
are not commercially available. A small number of forecast vendors
\emph{do} produce probabilistic forecasts but the calibration
methods they use are flawed. A number of academic papers have
suggested methods by which such forecasts could be improved but
again the methods described are flawed. To attempt to remedy this
situation we run a program of research aimed at clarifying the
issues involved in the creation of probabilistic temperature
forecasts and at developing methods that can be used to produce
such forecasts. We are not forecasters ourselves: our hope is that
the forecasting community will use the methods we describe to
produce forecasts that we can then use in our industrial
applications.

This article discusses a new method with the name Bayesian model
averaging (BMA) that has recently been proposed for the
calibration of temperature forecast ensembles
(see~\citet{raftery}). Our purpose is twofold:
\begin{enumerate}
    \item To point out the close connections between BMA and earlier
    methods known as kernel regression (KR) and kernel spread
    regression (KSR)
    \item To describe a number of flaws that we believe that the BMA approach suffers from
    that render it inappropriate as a method to be used for the calibration of real forecast data
\end{enumerate}

We start by describing the KR and BMA approaches. We then compare
the two and point out the problems we see in BMA. Finally we
suggest some further methods that take features from both BMA and
KR that could be used to solve the calibration problem that is
discussed in~\citet{raftery}.

\section{Kernel Regression}

Kernel regression (KR) was described by us in~\citet{jewson03h}.
It is a method that takes an ensemble forecast and turns it into a probabilistic forecast.
The simplest reasonable way to do this is to use linear regression on the ensemble mean.
KR is a simple extension of linear regression that allows for the representation of
non-normality in the temperature distribution by putting a small kernel
of optimum width around each ensemble member.
The probability density forecast from KR can be written as:

\begin{equation}\label{krp}
 p(x)=\sum_{i=1}^{M} p_i(x)
\end{equation}
where the $p_i$ are the individual kernels given by
\begin{equation}
    p_i(x) \sim N(x_i,\lambda^2)
\end{equation}
where $x_i$ is the $i$'th ensemble member and $\lambda$ is the bandwidth
 (these equations come from equation 1 in~\citet{jewson03h}).

In addition to applying kernels in this way the mean and the variance of the ensemble
members are calibrated using linear regression. We write the complete model as:
\begin{equation}
T_i \sim K (\alpha+\beta m_i, \gamma, \lambda)
\end{equation}

KR calibrates the ensemble mean using linear regression (which gives an optimal
combination between the ensemble mean and climatology) and fixes the spread and the non-normality
using the parameters $\gamma$ and $\lambda$.
The parameter $\lambda$ is the bandwidth of the kernels used and controls the smoothness
of the final predicted distribution. Small values of $\lambda$ lead to a multimodal distribution
while large values of $\lambda$ lead to a unimodal smooth distribution.

The mean of the prediction from KR is given by:
\begin{equation}\label{krm}
 E(x)=\alpha+\beta m_i
\end{equation}

while the variance of the prediction, which is constant in time
for the anomalies, is given by:

\begin{eqnarray}\label{krv}
 var(x)&=&\lambda^2+\frac{1}{M} \sum_{i=1}^{M} (x_i-\mu)^2\\\nonumber
       &=&\lambda^2+\gamma^2
\end{eqnarray}
or
\begin{equation}
\mbox{variance of modelled temperatures}=\lambda^2+\mbox{sample variance of calibrated ensemble members}
\end{equation}
(this equation is equation 9 in~\citet{jewson03h}).

An extension of KR that allows for the uncertainty to vary in time according
to variations in the ensemble spread is also described~\citet{jewson03h}, and can be written as

\begin{equation}
T_i \sim K (\alpha+\beta m_i, \gamma+\delta s_i, \lambda)
\end{equation}

This model, known as kernel spread regression (KSR), calibrates the ensemble spread
by having separate parameters for the mean and the variance of the spread. This was shown
to be necessary in~\citet{jewsonbz03a}.

The predicted variance from KSR is:
\begin{eqnarray}\label{ksrv}
 var(x)&=&\lambda^2+\frac{1}{M} \sum_{i=1}^{M} (x_i-\mu)^2\\\nonumber
       &=&\lambda^2+(\gamma+\delta s_i)^2
\end{eqnarray}

\section{Bayesian model averaging}

BMA is a general approach for combining the results from several statistical models using weights~\citep{hoeting}.
There are a number of ways that BMA could be used in the creation of probabilistic forecasts.
We will discuss the particular application of BMA given in~\citet{raftery}.
The conclusions we will draw do not apply to BMA in general, but only to this particular way of using
BMA.

The suggestion in~\citet{raftery} is that the probability density of future temperatures can be modelled as a weighted
sum of a number of probability densities from different sources:
\begin{equation}\label{bmap}
 p(x)=\sum_{i=1}^{i=M} w_i g_i(x)
\end{equation}

where
\begin{equation}
 g_i(x) \sim N (x_i,\sigma_i^2)
\end{equation}
where $x_i$ are the ensemble members
(these equations are equations 2 and 3 from~\citet{raftery}, written in our notation).

The variance of the probabilistic forecast is then given by
\begin{equation}
var(x)=\sum_{i=1}^{M} w_i (x_i-\mu)^2+\sum_{i=1}^{M} w_i \sigma_i^2
\end{equation}
(this is equation 7 from~\citet{raftery}).

\section{The connection between BMA and KR}

We now consider how BMA and KR are related.
To see the connection we consider a case where the individual forecasts are statistically
identical. BMA also considers the more general case where the forecasts are
statistically different although we will argue that since it
doesn't work in the simplest case of identical members it certainly can't be expected to work
in the more complex cases.

If the forecasts are statistically identical then
we can assume that the BMA weights and $\sigma_i$'s are equal:

\begin{eqnarray}
w_i&=&\frac{1}{M}\\
\sigma_i&=&\sigma
\end{eqnarray}

Equation~\ref{bmap} now gives:

\begin{equation}
 p(x)=\sum_{i=1}^{i=M} \frac{1}{M} g_i(x)
\end{equation}

and we can see that this agrees with equation~\ref{krp} if we define $g_i(x)=M p_i(x)$ i{.}e{.} if
we normalise the kernels differently. So this part of the two models is the same up to a simple
definition of the normalisation.

The BMA predicted mean is just
\begin{equation}\label{bmam}
 E(x)=\sum_{i=1}^{M} x_i
\end{equation}

i{.}e{.} the ensemble mean, and the BMA predicted variance is
\begin{equation}\label{bmav}
var(x)=\sum_{i=1}^{M} \frac{1}{M} (x_i-\mu)^2+\sigma^2
\end{equation}

We can now see the similarities and differences between BMA and kernel regression very clearly.
\begin{enumerate}

    \item By comparing equation~\ref{krm} with equation~\ref{bmam} we see that
    BMA predicts the expected temperature using the ensemble mean while KR predicts
    the expected temperature using an optimum combination of the ensemble mean with climatology

    \item By comparing equations~\ref{krv} and~\ref{ksrv} with equation~\ref{bmav} we see that
    BMA calibrates the mean level of uncertainty, the variability of the uncertainty
    and the smoothness of the distribution using a single parameter $\sigma$. KR uses two parameters to calibrate the
    mean level of uncertainty and the smoothness while KSR uses three parameters to calibrate
    the mean level of uncertainty, the variability of the uncertainty and the smoothness.

\end{enumerate}

BMA (when applied to the identical members case)
is a special case of KSR in which $\beta=1, \gamma=0$ and $\delta=1$.

\section{The problems with BMA}

Unfortunately Bayesian model averaging seems to suffer from a number of flaws
as a method for calibrating temperature ensembles.
These issues discussed below: the research on which these conclusions are
based is summarised in~\citet{jewson04l}.

The first problem concerns the calibration of the ensemble mean.
In the special case that we are considering
BMA predicts the expected temperature using the ensemble mean.
However it is well documented (\citet{leith}, \citet{vonstorch}, \citet{jewsonz03a}) that the
ensemble mean is not the optimal forecast for the expected temperature:
a `damped' version of the ensemble mean calculated using linear regression is better. This damping
performs an optimal calibration of the ensemble mean with climatology. An undamped ensemble
mean such as that produced by BMA does not have the correct variance and will not minimise RMSE.

The second problem concerns the calibration of the uncertainty.
To correctly calibrate the uncertainty of a probabilistic forecast one needs to consider (at least) two operations.
First, the temporal mean of the uncertainty must be fixed at an appropriate level. There is no information
about the temporal mean of the uncertainty in the ensemble itself: this information can only come from
past forecast error statistics. Secondly, the amplitude of the variability of the uncertainty must be
fixed at an appropriate level. Again, there is no information about the amplitude of the variability of
the uncertainty in the ensemble itself: this must be fitted from past forecast error statistics too.
What the ensemble provides is then the relative amplitude and phase of the fluctuations of the uncertainty.

The important point is that these two calibration steps
(calibrating the mean and the amplitude of the variability of the spread)
are independent. To set the mean level of the
uncertainty correctly one typically needs to inflate the ensemble spread. However, to set the amplitude
of the variations in the uncertainty correctly one may need to reduce the amplitude of the variations
in the ensemble spread. A statistical model thus needs at least two parameters in order
to calibrate spread correctly. If only one parameter is available, the calibration of the mean and the
variability of the uncertainty will be mixed together, and the results will be somewhat arbitrary and
very possibly less good than a calibration method that ignores the variability in the ensemble spread
altogether. This mixing of different aspects of the calibration
is what happens in BMA\footnote{and to be fair we should note that this problem also
arises in other forecast calibration methods that have been
suggested in the academic literature such as the methods of ~\citet{roulstons03} and~\citet{mylne02a}}.

KR, KSR and BMA add another operation in the calibration of the ensemble,
which is the smoothing of the ensemble towards or away from a normal distribution.
If the bandwidth of the kernel ($\lambda$ in kernel regression and $\sigma$ in BMA) is large then the ensemble
is smoothed towards a normal while if the bandwidth is small the probability forecast will likely be rather multimodal
and will have a shape that depends more strongly on the distribution of the individual ensemble members.
This smoothing operation needs a separate parameter to be performed correctly as it is an independent
issue from the calibration of the uncertainty.
KR and KSR use a separate parameter for this step while BMA
uses the same parameter as is used to calibrate the uncertainty.

In summary BMA only has a single free parameter ($\sigma$) rather than the three that are required to
perform the calibration that is being attempted. Thus the three operations that are being performed
(calibration of the mean level of the uncertainty, calibration of the variability of the uncertainty and
calibration of the smoothness of the forecast distribution) are mixed together. It is easy to
imagine situations in which this would cause problems.
For instance it would not be possible
for BMA to correctly calibrate an ensemble for which the variability in the ensemble spread contains
very little information (requiring a large value of $\sigma$) but in which the temporal mean of the ensemble
spread is close to the correct level (requiring a small value of $\sigma$). Nor would it be possible
for BMA to correctly calibrate an ensemble for which the ensemble spread was larger than the actual
uncertainty.

\section{The solution}

The solution to this problem is to use the correct number of free parameters for the calibration that is
being attempted.
Given only a single parameter the most sensible course of action seems to be to assume a normal distribution,
ignore the variations in spread and use the parameter to represent the mean level of uncertainty.
Given two parameters one should calibrate the mean and variability of the uncertainty, while still assuming a
normal distribution.
Finally given three parameters one can calibrate all three of the mean level of uncertainty, the
variability of the uncertainty and the smoothness.

\section{Weighted kernel regression}

In~\citet{raftery} BMA was used to combine a number of forecasts that were not statistically
identical. We have argued that BMA does not calibrate correctly in the statistically identical case,
and so cannot be expected to work in more general cases either.
How, then, should the original calibration problem described in~\citet{raftery} be solved?
The kernel regression models should not be used as is since they assume that
the forecasts are statistically identical.

One can imagine methods that take the best of the KSR and BMA approaches that might
include one or more of the following features:
\begin{itemize}
    \item the mean is predicted using multiple linear regression on the anomalies
    \item kernels with different widths are used on each ensemble member
    \item the kernels could be combined with different weights
    \item the uncertainty is predicted using some linear function on the weight ensemble spread
\end{itemize}

%The most obvious solution seems to be to generalise the kernel regression or kernel spread
%regression models to have variable weights, one for each input.
%We write these models as:
%
%\begin{equation}
%T_i \sim K (\alpha+\beta m_i, \gamma, (w_1, w_2, ..., w_M),\lambda)
%\end{equation}
%
%for weighted kernel regression
%and
%
%\begin{equation}
%T_i \sim K (\alpha+\beta m_i, \gamma+\delta s_i, (w_1, w_2, ..., w_M),\lambda)
%\end{equation}
%
%for weighted kernel spread regression.
%Presumably these models can be fitted rather easily using standard ML methods.

However, our previous experience of calibration suggests to us that much simpler models
might perform just as well since the effects of non-normality and the benefit of using
the spread may well both be small. In that case multiple linear regression on the anomalies
is probably ideal, and whatever method is being used it should be compared with
linear regression on the anomalies as an appropriate minimal model.

\section{Summary}

We have discussed the question of how to produce probabilistic forecasts of temperature.
In particular we have dissected the Bayesian model averaging approach of~\citet{raftery}.
This approach is very similar to an earlier approach known as kernel regression~\citep{jewson03h}.
We have argued that BMA does not calibrate temperatures in an appropriate way.
Neither the predicted mean nor the predicted variance are constructed accurately.
With respect to the predicted mean, the issue of `damping' towards climatology has been omitted.
With respect to the variance, BMA mixes the separate functions of calibrating the mean level of
uncertainty, the amplitude of the variability of the uncertainty
and the smoothness of the forecast distribution into a single factor.
We conclude that BMA (as applied in~\citet{raftery}) is not a calibration method at all,
but simply a method to fit a distribution to a set
of ensemble members. As such it is more or less the same
as the well known kernel density of classical statistics.

\section{Acknowledgements}

Thanks to Christine Ziehmann for interesting discussions on this topic.

\section{Legal statement}

SJ was employed by RMS at the time that this article was written.

However, neither the research behind this article nor the writing
of this article were in the course of his employment, (where 'in
the course of their employment' is within the meaning of the
Copyright, Designs and Patents Act 1988, Section 11), nor were
they in the course of his normal duties, or in the course of
duties falling outside his normal duties but specifically assigned
to him (where 'in the course of his normal duties' and 'in the
course of duties falling outside his normal duties' are within the
meanings of the Patents Act 1977, Section 39). Furthermore the
article does not contain any proprietary information or trade
secrets of RMS. As a result, the author is the owner of all the
intellectual property rights (including, but not limited to,
copyright, moral rights, design rights and rights to inventions)
associated with and arising from this article. The author reserves
all these rights. No-one may reproduce, store or transmit, in any
form or by any means, any part of this article without the
author's prior written permission. The moral rights of the author
have been asserted.

The contents of this article reflect the author's personal
opinions at the point in time at which this article was submitted
for publication. However, by the very nature of ongoing research,
they do not necessarily reflect the author's current opinions. In
addition, they do not necessarily reflect the opinions of the
author's employer.

\bibliography{jewson}

\end{document}